\documentclass[12pt]{article}
\usepackage{graphicx, here, amsfonts, epsfig, hhline}
\newcommand{\bea}{\begin{eqnarray}}
\newcommand{\eea}{\end{eqnarray}}

\begin{document}

\begin{titlepage}
\hskip 11cm \vbox{ \hbox{DFCAL-TH 04/2} \hbox{April 2004}} \vskip 1cm

\phantom{.} \vspace{-1cm} \centerline{\Large \bf  Baryonic Regge
trajectories with analyticity constraints}

\vskip 2cm

\centerline{R.~Fiore$^{a\dagger}$, L. L.~Jenkovszky$^{b\ddagger}$,
F.~Paccanoni$^{c\ast}$, A.~Prokudin$^{d\diamond}$}

\vskip 0.1cm

\centerline{$^{a}$ \sl Dipartimento di Fisica, Universit\`a della Calabria and}
\centerline{\sl Instituto Nazionale di Fisica Nucleare, Gruppo collegato di Cosenza}
\centerline{\sl I-87036 Arcavata di Rende, Cosenza, Italy}

\centerline{$^{b}$ \sl Bogolyubov Institute for Theoretical Physics}
\centerline{\sl Academy of Science of Ukraine}
\centerline{\sl UA-03143 Kiev, Ukraine}

\centerline{$^{c}$ \sl Dipartimento di Fisica, Universit\`a di Padova and}
\centerline{\sl Instituto Nazionale di Fisica Nucleare, Sezione di Padova}
\centerline{\sl Via F. Marzolo 8, I-35131 Padova, Italy}

\centerline{$^{d}$ \sl  Dipartimento di Fisica Teorica, Universit\`a di
Torino,}
\centerline{\sl Instituto Nazionale di Fisica Nucleare, Sezione di Torino}
\centerline{\sl Via P. Giuria 1, I-10125 Torino, Italy and}
\centerline{\sl Institute for High Energy Physics, 142284 Protvino, Russia} 

\vskip 1cm

A model for baryonic Regge trajectories compatible with the
threshold behavior required by unitarity and asymptotic behavior in
agreement with analyticity constraints is given in explicit form.
Widths and masses of the baryonic resonances on the $N$ and
$\Delta$ trajectories are reproduced. The MacDowell symmetry is
exploited and an application is given.
\vfill

\hrule
$
\begin{array}{ll}
^{\dagger}\mbox{{\it e-mail address:}} &
  \mbox{FIORE@CS.INFN.IT} \\
^{\ddagger}\mbox{{\it e-mail address:}} &
  \mbox{JENK@GLUK.ORG} \\
^{\ast}\mbox{{\it e-mail address:}} &
  \mbox{PACCANONI@PD.INFN.IT} \\
^{\diamond}\mbox{{\it e-mail address:}} &
  \mbox{PROKUDIN@TO.INFN.IT} 
\end{array}
$
\end{titlepage}
\eject
\newpage

\section{Introduction}

The dynamical origin of the meson mass spectra and the analytic
properties of the bosonic Regge trajectories have been studied
thoroughly in the past and a comprehensive review on this subject
can be found in Ref.~\cite{AEI}. The same statement does not hold for
baryons that present further complications due to their more
involved internal structure and, as we shall see, to the intricate
analytical properties of their Regge trajectories. \vskip 0.3cm
Information on the $N$ and $\Delta$ baryon trajectories has been
obtained from the analysis of backward pion-nucleon scattering,
see for example, Refs.~ \cite{BC,CS,GKT,KN}, where the important problem
of the linearity of the baryon Regge trajectories has been
investigated in detail. The conclusion that Regge trajectories are
not straight and parallel lines is supported by the detailed
analysis of Ref.~\cite{ATN}. Moreover, in the case of baryons, the
analyticity properties of the trajectory function \cite{VNG}, that
follow from the invariance under Schwinger's space-time reflection
of the covariant scattering amplitude \cite{SMD}, confirm this
conclusion. \vskip 0.3cm Since the position of the singularity in
the $J$ plane of the partial wave amplitude is an analytic
function of the center of mass energy in the relevant channel, say
$\sqrt{s}$, the MacDowell symmetry \cite{SMD} implies that
\begin{equation}
\alpha^+(\sqrt{s})=\alpha^-(-\sqrt{s})\;\;\;\mbox{for}\;\;s>0~,
\label{T0}
\end{equation}
where $\pm$ denote the parity of the trajectory. The relation
(\ref{T0}) requires that, if the trajectory is linear in $s$,
parity doublets must exist. The simplest way of eliminating
doublets, that are not observed experimentally, is to take into
account deviations of the trajectories from linearity and to show, as
in Ref.~\cite{KN}, that these deviations are compatible with
experimental data. Arguments based on the spontaneous chiral
symmetry breaking in the low energy part of the baryon spectrum
\cite{LYG} support this point of view. \vskip 0.3cm For the above
reasons, attempts to exploit the explicit form of the baryon
trajectories on the basis of the experimental data only meet great
difficulties. While in Ref.~\cite{GKT} the conclusion is that
$N_{\alpha}$ and $N_{\beta}$ are independent Regge trajectories,
since otherwise it would be impossible to explain the energy
dependence of $\pi^+\,p$ backward cone and the dip, the
parametrization of Ref.~\cite{KN} succeeds in reproducing the baryon
spectrum, the energy dependence of cross sections and the
momentum transfer dependence of differential cross sections.
Dispersion relations for the trajectory function
\cite{GKT,VNG,PDBC} impose severe constraints on the analytic
structure of this function \cite{ADP,FPS} and give the opportunity
to restrict its possible form. \vskip 0.3cm In this paper we
construct an explicit model for complex Regge trajectories
reproducing both the masses and widths of observed baryonic
resonances with the constraints of analyticity and unitarity.
Section {\bf 2} presents an attempt to adapt a previous model for
meson trajectories \cite{RAL,RAAL} to the baryon spectrum. The $N$
and $\Delta$ trajectories are considered in detail in Section {\bf
3}. In Section {\bf 4} we implement the MacDowell symmetry and
derive explicit formulas for the real and imaginary parts of the
trajectory. The application to the nucleon trajectory is studied
in Section {\bf 5}. The last Section is devoted to concluding remarks.

\section{A simple model}

The properties of the bosonic trajectories following from
analyticity and unitarity \cite{BZ} have been summarized in our
previous papers \cite{RAL,RAAL}. Fermion trajectories suffer
further complications \cite{PDBC}. The generalization of the
MacDowell symmetry shows that, in order to satisfy the relation
between natural and unnatural parity amplitudes, we need two
trajectories of opposite parity that are related by the relation 
(\ref{T0}). Moreover the dispersion relation for the trajectory
function should exhibit analyticity in $\sqrt{s}$ and should be
written in terms of this variable. In order to clarify the
problem, however, we will limit ourselves, in this Section, to
consider analyticity in the variable $s$ for a simplified model.
\vskip 0.3cm Let $P$ be the parity of the resonances lying on a
Regge trajectory. Since $P=\eta (-1)^{J-v}$, where $v=1/2$ for odd
half-integral $J$ and $v=0$ for integral $J$, natural parity means
$\eta=+1$ while unnatural parity means $\eta=-1$. Hence for the $N$
trajectory, $(\frac{1}{2}^+,\frac{5}{2}^+,\;\ldots)$, we have
$\eta=+1$ while, for the $\Delta$ trajectory,
$(\frac{3}{2}^+,\frac{7}{2}^+,\;\ldots)$, $\eta=-1$. The minimum
allowed angular momentum is \cite{BZ}
\begin{displaymath}
L=J-(s_1+s_2)+\frac{1}{2}\left[
1-\eta\eta_1\eta_2\,(-1)^{s_1+s_2-v}\right]
\end{displaymath}
and, since $\eta_1=-1$ for the pion and $\eta_2=+1$ for the
nucleon, the $\Delta$ trajectory function will be characterized by
$L=J-\frac{1}{2}$. Both the real and imaginary part of the
trajectory will inherit the threshold behavior of the partial wave
amplitude for pion-nucleon scattering:
\begin{equation}
(q_{12}^2)^{L+1/2}=(q_{12}^2)^J \; ,\label{b1}
\end{equation}
where $q_{12}$ is the center of mass momentum. \vskip 0.3cm While
retaining the assumption of additivity of threshold contributions,
the more regular behavior of the baryonic resonance widths suggest that the
choice of the imaginary part of the trajectory can be different
from the bosonic case. As a first attempt, we consider analyticity
in $s$ for the trajectory functions and start from the simple form
for the imaginary part of the trajectory
\begin{equation}
{\cal I}m \alpha(s)=s^{\delta} \sum_n c_n
\left(\frac{s-s_n}{s}\right)^{{\cal R}e \alpha(s_n)} \cdot
\theta(s-s_n)~.  \label{b2}
\end{equation}
Eq.~(\ref{b2}) has the correct threshold behaviour and
analyticity requires that $\delta <1$. The boundedness of
$\alpha(s)$ for $s \to \infty$ follows from the condition that the
amplitude, in the Regge form, should have no essential singularity
at infinity in the cut plane. \vskip 0.3cm The once subtracted
dispersion relation for the trajectory is
\begin{equation}
{\cal R}e\,\alpha(s)=\alpha(0)+\frac{s}{\pi}\,PV\int_0^{\infty}
ds' \frac{{\cal I}m \alpha(s')}{s'(s'-s)}~, \label{b0}
\end{equation}
where $PV$ means the Cauchy principal value of the integral.
Setting $\lambda_n={\cal R}e \alpha(s_n)$ we get
\begin{equation}
{\cal R}e\,\alpha(s)=\alpha(0)+\frac{s}{\pi}\sum_n c_n {\cal
A}_n(s)~, \label{b3}
\end{equation}
where~\cite{BAT}
\begin{eqnarray*}
& & {\cal A}_n(s)=
\frac{\Gamma(1-\delta)\Gamma(\lambda_n+1)}{\Gamma(\lambda_n-\delta+2)
s_n^{1-\delta}}{}_2F_1\left(1,1-\delta;\lambda_n-\delta+2;\frac{s}{s_n}\right)\theta(s_n-s)+\\
& & \left\{ \pi s^{\delta-1}\left(
\frac{s-s_n}{s}\right)^{\lambda_n} \cot[\pi(1-\delta)]- \right. \\
& & \left.
\frac{\Gamma(-\delta)\Gamma(\lambda_n+1)s_n^{\delta}}{s\Gamma(\lambda_n-\delta+1)
}{}_2F_1\left(\delta-\lambda_n,1;\delta+1;\frac{s_n}{s}\right)
\vphantom{\left(
\frac{s-s_n}{s}\right)^{\lambda_n}} \right \} \theta(s-s_n)~.
\end{eqnarray*}
\vskip 0.3cm From these equations we get the slope
\begin{equation}
{\cal R}e\,\alpha'(s)= \frac{1}{\pi}\sum_n c_n {\cal B}_n(s)~,
\label{b4}
\end{equation}
where
\begin{eqnarray*}
& & {\cal B}_n(s)=
\frac{\Gamma(1-\delta)\Gamma(\lambda_n+1)}{\Gamma(\lambda_n-\delta+2)
s_n^{1-\delta}}{}_2F_1\left(2,1-\delta;\lambda_n-\delta+2;\frac{s}{s_n}\right)\theta(s_n-s)+\\
& & \left\{ \pi s^{\delta-1}\left(
\frac{s-s_n}{s}\right)^{\lambda_n} 
\cot[\pi(1-\delta)]\left(\delta+\lambda_n\frac{s_n}{s-s_n}\right)- \right. \\
& & \left.
\frac{\Gamma(-\delta)\Gamma(\lambda_n+1)s_n^{\delta+1}}{(1+d)\Gamma(\lambda_n-\delta)s^2
}{}_2F_1\left(1+\delta-\lambda_n,2;\delta+2;\frac{s_n}{s}\right)
\vphantom{\left(
\frac{s-s_n}{s}\right)^{\lambda_n}}
\right\} \theta(s-s_n)~.
\end{eqnarray*}
Eqs.~(\ref{b2}) and (\ref{b4}) determine the widths of the
resonances through the relation
\begin{equation}
\Gamma=\frac{{\cal I}m\,\alpha(M^2)}{M{\cal R}e\,\alpha'(M^2)}~,
\label{b5}
\end{equation}
where $\Gamma$ is the total width of the resonance and $M$ is its
mass.

\section{Fitting the resonances masses and widths in the
simplified model}

In this section we apply the formalism to the $\Delta$ baryon
trajectory $(\frac{3}{2}^+,\frac{7}{2}^+,\;\ldots)$ and  nucleon
$N^+$ trajectory which contains
baryons N(939) $\frac{1}{2}^+$, N(1680) $\frac{5}{2}^+$, N(2220) 
$\frac{9}{2}^+$ and N(2700) $\frac{13}{2}^+$ \cite{PDG}. In the fit the input 
data are the masses and widths of the resonances. The quantities to be
determined are the parameters $c_n$, $\delta$ and the thresholds
$s_n$. While $s_1$ is fixed to the pion-nucleon threshold, data allow for 
the determination of other two thresholds at maximum. i.e. $s_2$ and a
higher threshold, the one above all the  known resonances, that 
will be called  $s_x$ in the following.

\subsection{$\Delta^+$ trajectory}

First we fit the real part of the amplitude using a linear
form of the trajectory
\begin{equation}
{\cal R}e\,\alpha(s)=\alpha(0)+\alpha'(0)\cdot s\; .
\end{equation}
Thus we obtain $\alpha(0)=0.17\pm 0.02$ and 
$\alpha'(0)=0.87\pm 0.01$ (GeV$^{-2}$) ($\chi^2/$d.o.f = 0.63).
Using these values of the parameters we calculate ${\cal R}e\,\alpha(s_n)$,
where $n=1,2,x$ and $s_1=(m_{\pi}+m_N)^2=1.16$ GeV$^2$,
$s_2$ = 2.8 GeV$^2$ and $s_x$ = 15 GeV$^2$.

\noindent
Then, as previously stated, we set $\lambda_n={\cal R}e\,\alpha(s_n)$ and continue 
the recursive fitting procedure using formulas (\ref{b3}) and (\ref{b4}).
After 15 steps the procedure converges and we obtain the values
of the parameters:
$\alpha(0)=-0.04\pm 0.03$, $\delta = -0.26 \pm 0.01$, $c_1=0.75\pm 0.03$, 
$c_2=2.2\pm 0.3$ and $c_x=1414 \pm 75$, $s_2 = 2.44 \pm 0.05$ GeV$^2$,
$s_x = 11.7 \pm 0.3$ GeV$^2$ ($\chi^2/$d.o.f = 1.4).

$\Delta $ resonances have been seen in a large number of formation and 
production experiments. $\Delta(1950)$ has been seen in the $\Sigma K$ 
channel, $\pi^+p\to \Sigma^+K^+$ , but the evidence is poor, and one 
of its decay modes is $N\rho$ , $(<10\%)$. It is interesting to notice 
that the threshold in the $\Sigma K$ channel, for example, is $2.84$  
GeV${}^2$ and for $N\rho$ is $2.92$ GeV${}^2$. Both are higher than, 
but not too far from, the second threshold $s_2=2.44$ GeV${}^2$ found 
empirically minimizing the $\chi^2$ of the fit. 

\begin{figure}[H]
\parbox[r]{8.5cm}{\epsfxsize=75mm
\epsffile{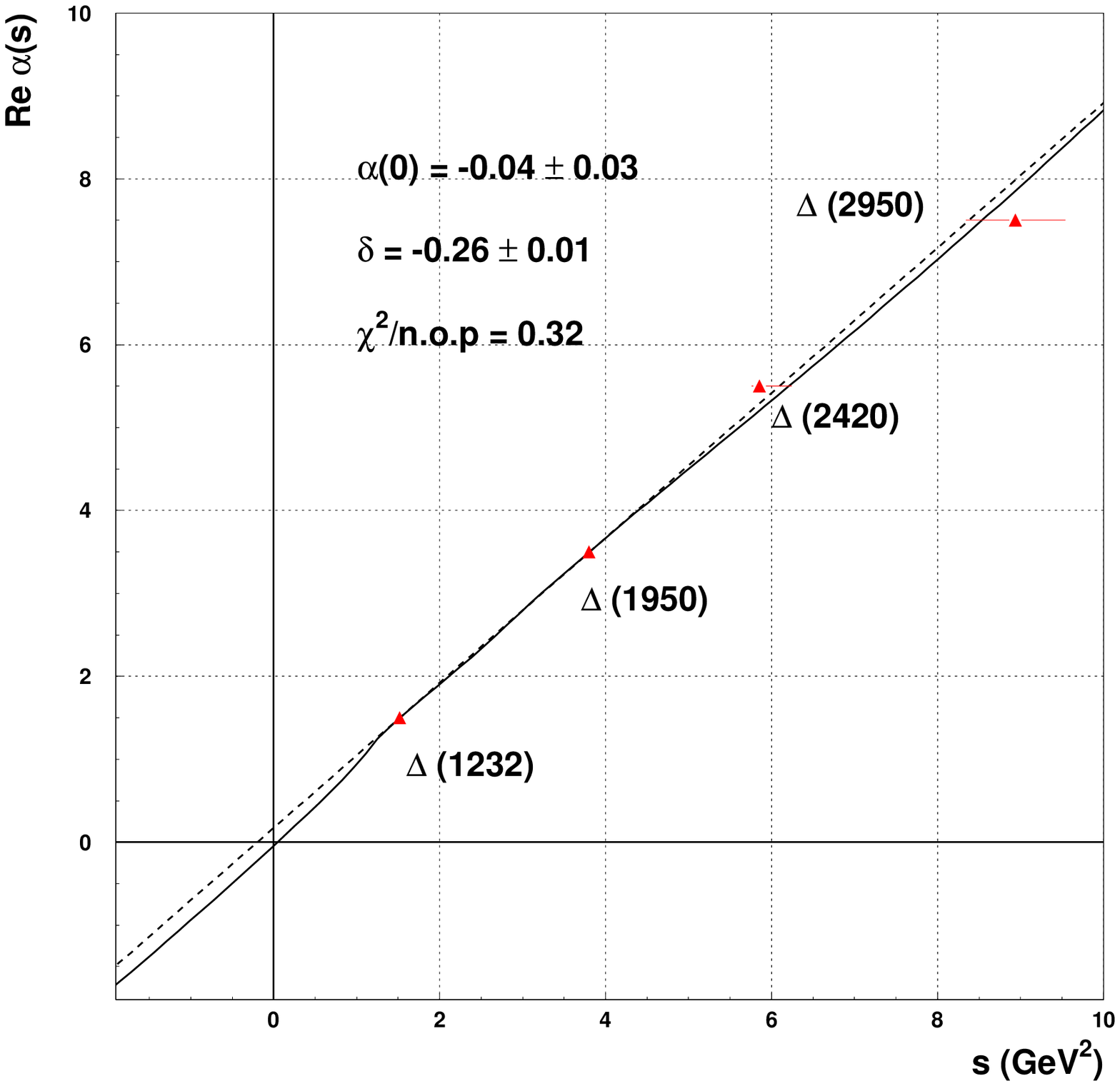}} \hfill~\parbox[c]{9.cm}{\epsfxsize=75mm
\epsffile{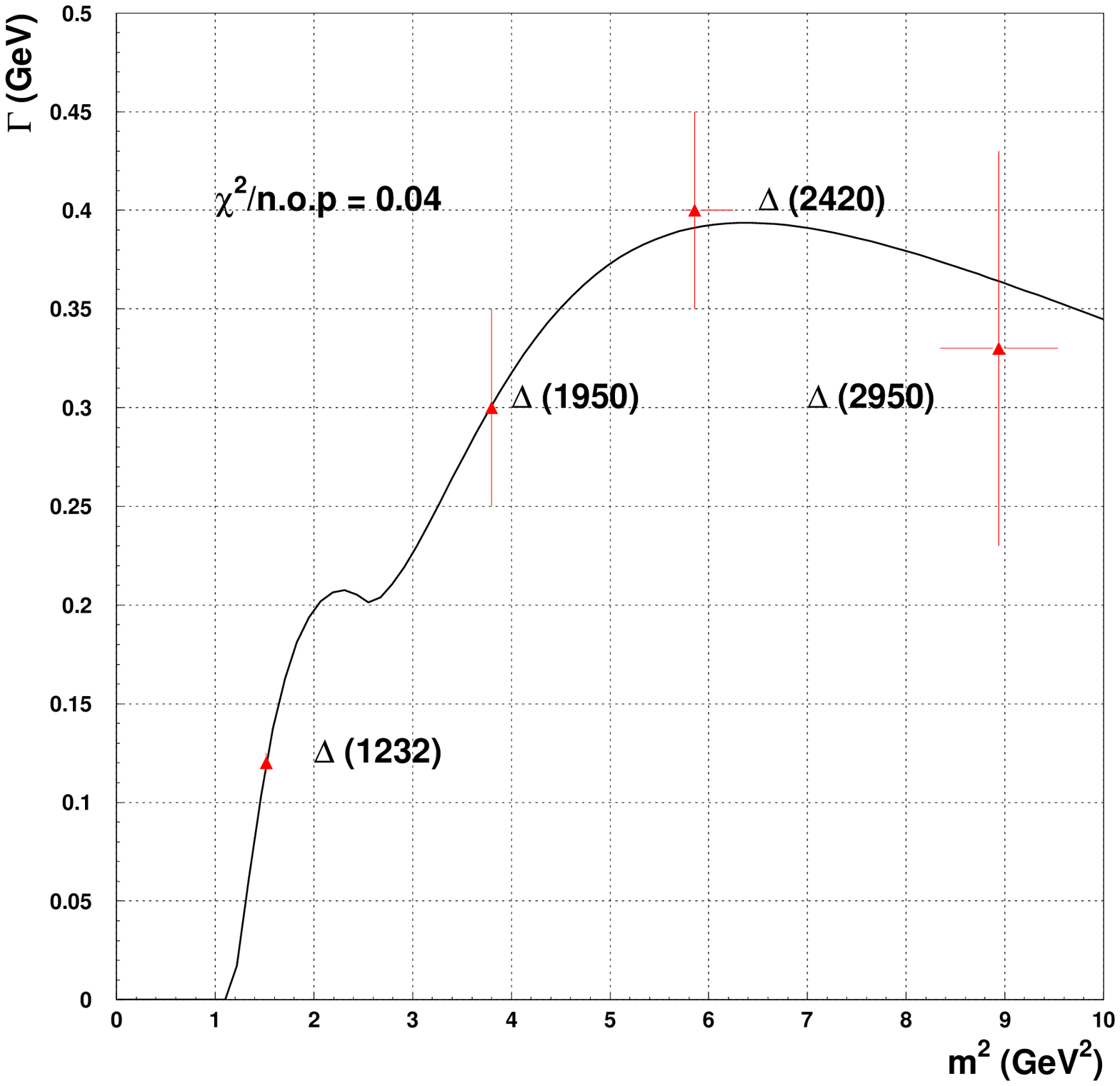}}

\parbox[t]{8.cm}{\caption{The real part of $\Delta$ trajectory. The dashed line
corresponds to the result of a linear fit, the solid line corresponds to the final result.
\label{fig:delta}}}
\hfill~\parbox[t]{8.cm}{\caption{ The width of $\Delta$ trajectory.
\label{fig:deltawidth}}}
\end{figure}

\noindent
The results are shown in figures~\ref{fig:delta} and \ref{fig:deltawidth}. 

\subsection{$N^+$ trajectory}

As in the previous Subsection, we start fitting the real part of the 
amplitude using a linear form of the trajectory:
\begin{equation}
{\cal R}e\,\alpha(s)=\alpha(0)+\alpha'(0)\cdot s\; .
\end{equation}
Thus we obtain $\alpha(0)=-0.27\pm 0.14$ and 
$\alpha'(0)=0.98\pm 0.04$ (GeV$^{-2}$) ($\chi^2/$d.o.f = 0.24).
Using these values of the parameters we calculate ${\cal R}e\,\alpha(s_n)$,
where $n=1,2,x$ and $s_1=(m_{\pi}+m_N)^2=1.16$ GeV$^2$,
$s_2$ = 2.44 GeV$^2$ and $s_x$ = 11.7 GeV$^2$. Now the channels $N\eta$, 
with a threshold at 2.21 GeV$^2$, and $\Lambda K$, with a threshold at 
2.59 GeV$^2$, are no more forbidden. Also in the $N^+$ the interpretation 
of the second threshold as a physical one seems now appropriate.

\noindent
Then we set $\lambda_n={\cal R}e\,\alpha(s_n)$ and continue the 
recursive fitting procedure using formulas (\ref{b3}) and (\ref{b4}).
Again after 15 steps the procedure converges and we obtain the values
of the parameters:
$\alpha(0)=-0.41$, $\delta = -0.46 \pm 0.07$, $c_1=0.51\pm 0.08$, 
$c_2=4.0\pm 0.8$ and $c_x=(4.5 \pm 1.7)\cdot 10^{+4}$ ($\chi^2/$d.o.f = 1.15).

\begin{figure}[H]
\parbox[r]{8.5cm}{\epsfxsize=75mm
\epsffile{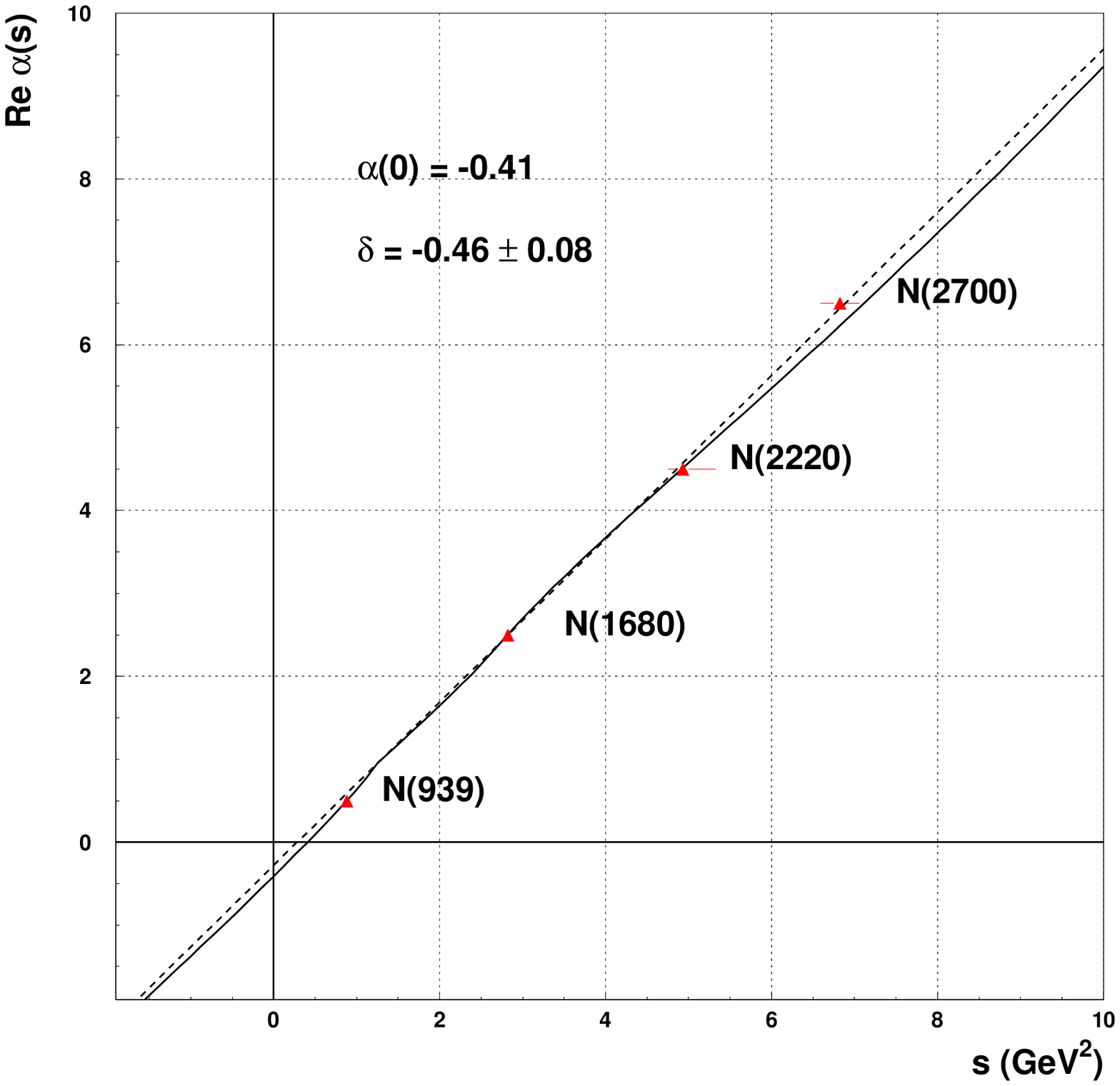}} \hfill~\parbox[c]{9.cm}{\epsfxsize=75mm
\epsffile{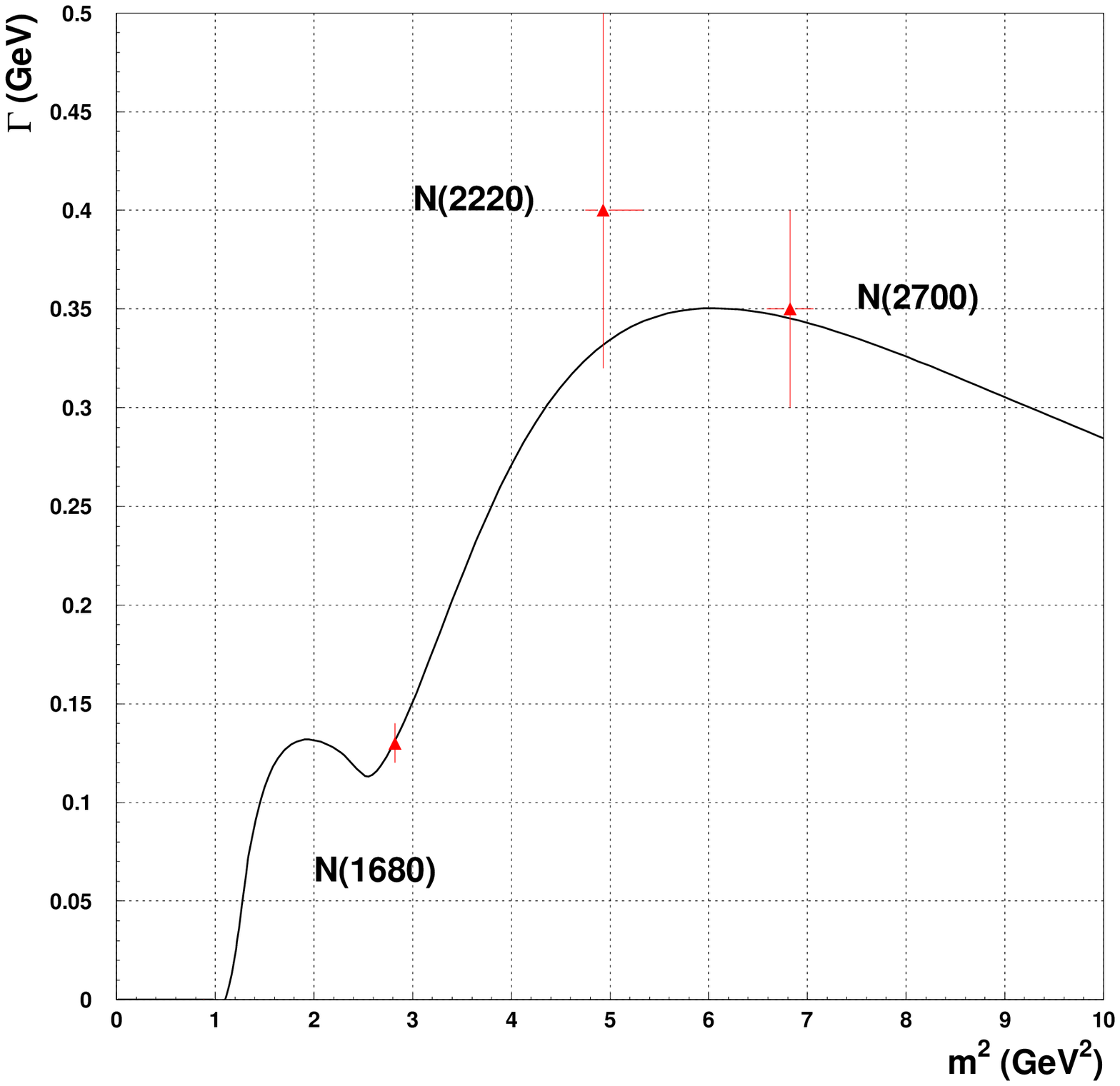}}

\parbox[t]{8.cm}{\caption{The real part of $N$ trajectory. The dashed line
corresponds to the result of a linear fit, the solid line corresponds to the final result.
\label{fig:n1}}}
\hfill~\parbox[t]{8.cm}{\caption{ The width of $N$ trajectory.
\label{fig:n1width}}}
\end{figure}

\noindent
The results are depicted in figs.~\ref{fig:n1} and \ref{fig:n1width}. 

\section{An approach based on MacDowell symmetry}

The trajectories of baryonic Regge poles in states with angular
momentum $J$ and parity $(-1)^{J\pm 1/2}$ coincide when the c.m.
energy squared $s$ tends to zero and become complex conjugate of
each other for $s<0$ \cite{VNG}. This relation is due to the
kinematic singularity $\sqrt{s}$ present in spinor amplitudes. The
trajectory function $\alpha=\alpha(\sqrt{s})$ is complex for
$\sqrt{s} > \sqrt{s_1}$ and $\sqrt{s} < -\sqrt{s_1}$ and real in
the interval $-\sqrt{s_1}<\sqrt{s}<\sqrt{s_1}$ where $s_1$ is the
first threshold, $s_1=(m_{\pi}+m_N)^2=1.16\;GeV^2$ for $\Delta$
and $N$ trajectories. A dispersion relation can be written for
$\alpha(\sqrt{s})$ \cite{VNG,BC,PDBC}:
\begin{equation}
\alpha(\sqrt{s})=\frac{1}{\pi}\int_{\sqrt{s_1}}^{\infty}\,\frac{
{\cal I}m\,\alpha^+(s')}{\sqrt{s'}-\sqrt{s}}\,d\sqrt{s'}+
\frac{1}{\pi}\int_{\sqrt{s_1}}^{\infty}\,\frac{ {\cal
I}m\,\alpha^-(s')}{\sqrt{s'}+\sqrt{s}}\,d\sqrt{s'}~, \label{M1}
\end{equation}
where ${\cal I}m\,\alpha^{\pm}(s)$ are the imaginary parts of the
position of the poles of the $\pi-N$ scattering amplitudes in
states with angular momentum $\alpha$ and parity $(-1)^{\alpha \pm
1/2}$. \vskip 0.3cm The MacDowell symmetry for the $\pi-N$ partial
waves amplitudes \cite{SMD}, 
\begin{equation}
T_{J-1/2}^J (\sqrt{s})=T_{J+1/2}^J (-\sqrt{s})~, \label{M2}
\end{equation}
implies the existence of resonance states with the same $J$, but
opposite parity, except when their residue vanishes. Hence the
possibility to reveal these additional resonances, and confirm
their existence, depends on the detailed behavior of the residue
function \cite{JAS,BC,CS}\,\footnote{ The absence of parity
doublets in the low energy part of baryon spectrum as an
indication of spontaneous symmetry breaking has been proposed in
Ref~\cite{LYG}}. \vskip 0.3cm In the following we will consider the
$N$-trajectory and express the $\ell=J+1/2$ amplitude through the
$\ell=J-1/2$ amplitude by using Eq.~(\ref{M2}). Then the $N$
trajectory, $\alpha_N(\sqrt{s})$, at the resonances will have the
form: $\alpha_N(-2.2)=9/2
,\;\alpha_N(-1.68)=5/2,\;\alpha_N(-0.939)=1/2,\;\alpha_N(1.675)=
5/2,\;\alpha_N(2.25)=9/2$ where the masses of the resonances have
been expressed in GeV. An updated fit to these resonances has been
performed with a trajectory that splits the degeneracy \cite{CS}:
\begin{equation}
\alpha_N(\sqrt{s})=a_0+a_1\sqrt{s}+a_2s~, \label{M3}
\end{equation}
where $a_0$ can be eliminated by imposing the condition
$\alpha_N(-0.939)=1/2$. The result of this fit, shown in figure
~\ref{fig:n}, is interesting and such as to justify a deeper analysis.
\vskip 0.3cm On the basis of the previous results for the
asymptotic behavior of ${\cal I}m\,\alpha(s)$ we perform only one
subtraction in Eq.~(\ref{M1}) and, for the real part of
$\alpha(\sqrt{s})$, we get
\begin{eqnarray}
{\cal R}e\,\alpha(\sqrt{s})=\alpha(0) & + &
\frac{\sqrt{s}}{\pi}\,PV\int_0^{\infty} \frac{ {\cal
I}m\,\alpha^+(s')}{\sqrt{s'}(\sqrt{s'}-\sqrt{s})}\,d\sqrt{s'} \nonumber \\
& - & \frac{\sqrt{s}}{\pi}\,PV\int_{0}^{\infty}\,\frac{ {\cal
I}m\,\alpha^-(s')}{\sqrt{s'}(\sqrt{s'}+\sqrt{s})}\,d\sqrt{s'}~,
\label{M4}
\end{eqnarray}
where only one of the two principal values has to be taken: the first
one for $\sqrt{s}>0$ and the second one for $\sqrt{s}<0$. It is
convenient to consider the equivalent form
\begin{eqnarray}
& & {\cal
R}e\,\alpha(\sqrt{s})=\alpha(0)+\frac{\sqrt{s}}{2\pi}\,PV\int_0^{\infty}
\frac{ds'}{s'(s'-s)} \left( \sqrt{s'} \left[{\cal
I}m\,\alpha^+(s')-{\cal I}m\,\alpha^-(s') \right] \right.
\nonumber
\\ & & +\left. \sqrt{s} \left[{\cal I}m\,\alpha^+(s')+{\cal
I}m\,\alpha^-(s')\right] \Big)~, \right. \label{M5}
\end{eqnarray}
where the simple expression (\ref{b0}), studied before, is
re-obtained when ${\cal I}m\,\alpha^+(s)={\cal I}m\,\alpha^-(s)$.
Obviously, the solution of Eqs.~(\ref{M4}) and (\ref{M5}) is
the same but the proof is not trivial and will be given in
Appendix A. \vskip 0.3cm The experimental values of the widths of
resonances, symmetric with respect to the axis $\sqrt{s}=0$ in the
$\sqrt{s} - J$ plane, coincide within the errors. Hence their
imaginary parts cannot be too different and a minor
modification of the previous assumption (\ref{b2}),
\begin{equation}
{\cal I}m\,\alpha^{\pm}(s)=s^{\delta}\sum_n c_n
\left(\frac{s-s_n}{s}\right)^{{\cal R}e \alpha^{\pm}(s_n)} \cdot
\theta(s-s_n)~,\label{M6}
\end{equation}
seems reasonable in view of the small difference  between the
trajectories
\begin{displaymath}
\alpha^{\pm}(\sqrt{s})=a_0\pm a_1\sqrt{s}+a_2 s
\end{displaymath}
resulting from the fit. The condition $\delta< 1/2$, necessary for
the convergence of integrals in Eq.~(\ref{M4}), has been
implemented in view of the results in the preceding Section. In
Eq.~(\ref{M6}) ${\cal I}m\,\alpha^{\pm}(s)$ does not change under
the transformation $\sqrt{s}\to -\sqrt{s}$ and hence, for example,
$s^{\delta}$ must be interpreted as $(|\sqrt{s}|^2)^{\delta}$
while $\sqrt{s_n}$ is always positive. \vskip 0.3cm Let us rewrite
Eq.~(\ref{M5}) in the form
\begin{equation}
{\cal R}e\,\alpha(\sqrt{s})=\alpha(0)+\sum_n c_n {\cal
F}_n(\sqrt{s}) \label{M7}
\end{equation}
and set $\lambda_n^{\pm}={\cal R}e\,\alpha^{\pm}(s_n)$. With this
input in Eq.~(\ref{M5}) the expression for ${\cal F}_n(\sqrt{s})$
can be easily evaluated. We find
\begin{eqnarray}
{\cal F}_n(\sqrt{s}) & = & \frac{1}{2} \left\{ \left[s^{\delta}
\left(1-\frac{s_n}{s}\right)^{\lambda_n^+}
\tan(\pi\delta)-\frac{\sqrt{s_n}}{\sqrt{s}} \frac{s_n^{\delta}
\Gamma(-1/2-\delta)\Gamma(1+\lambda_n^+)}{\pi
\Gamma(1/2-\delta+\lambda_n^+)}\times \right. \right. \nonumber \\
& \times & \left. {}_2F_1\left(1,\frac{1}{2}+\delta-\lambda_n^+;
\frac{3}{2}+\delta;\frac{s_n}{s}\right)-\mbox{the same term
with}\;\;\lambda_n^+ \to \lambda_n^-\right] + \nonumber \\
& + & \left[-s^{\delta} \left(1-\frac{s_n}{s}\right)^{\lambda_n^+}
\cot(\pi\delta)- \frac{s_n^{\delta}
\Gamma(-\delta)\Gamma(1+\lambda_n^+)}{\pi
\Gamma(1-\delta+\lambda_n^+)}{}_2F_1\left(1,\delta-\lambda_n^+;
1+\delta;\frac{s_n}{s}\right) +\right.\nonumber \\ & + & \left.
\left. \mbox{the same term with}\;\;\lambda_n^+ \to
\lambda_n^-
\vphantom{\left(1-\frac{s_n}{s}\right)^{\lambda_n^+}}\right ]\right \} \theta(s-s_n) \nonumber \\
& + & \frac{1}{2} \left\{ \left[\frac{\sqrt{s}}{\sqrt{s_n}}
\frac{s_n^{\delta} \Gamma(1/2-\delta)\Gamma(1+\lambda_n^+)}{\pi
\Gamma(3/2-\delta+\lambda_n^+)}\times \right. \right. \nonumber \\
& \times & \left. {}_2F_1\left(1,\frac{1}{2}-\delta;
\frac{3}{2}-\delta+\lambda_n^+;\frac{s}{s_n}\right)-\mbox{the same
term with}\;\;\lambda_n^+ \to \lambda_n^-\right] + \nonumber
\\ & + & \left[\frac{s}{s_n} \frac{s_n^{\delta}\Gamma(1-\delta)
\Gamma(1+\lambda_n^+)}{\pi\Gamma(2-\delta+\lambda_n^+)}
{}_2F_1\left(1,1-\delta;2-\delta+\lambda_n^+;\frac{s}{s_n}\right)
\right. \nonumber
\\ & + & \left. \left. \mbox{the same term with} \;\;\lambda_n^+
\to \lambda_n^- \vphantom{\left(1-\frac{s_n}{s}\right)^{\lambda_n^+}}\right ]\right \} \theta(s_n-s)~. \label{M8}
\end{eqnarray}
\vskip 0.3cm 
The widths can be found by first
calculating the derivative of $\alpha'{}_R(\sqrt{s})$. The result can be
written as
\begin{displaymath}
\alpha'{}_R(\sqrt{s}) \equiv \frac{d{\cal R}e
\alpha(\sqrt{s})}{d\sqrt{s}}=\sum_n c_n{\cal D}_n(\sqrt{s})~.
\end{displaymath}
By comparison with a Breit-Wigner resonance of mass $M$, where $M$ can
be positive or negative in this context, we get the total width
\begin{equation}
\Gamma=\frac{2 {\cal I}m\,\alpha(M^2)}{\left|
\alpha'{}_R(M)\right|}~.\label{M12}
\end{equation}
The calculation of $\alpha'{}_R(\sqrt{s})$ parallels the one
already done in the preceding Section and the explicit form of
${\cal D}_n(\sqrt{s})$ will be shown in Appendix B.

\section{Application to the nucleon trajectory}

In this section we apply the formalism to the nucleon trajectory. 

\noindent
First we fit the real part of the amplitude using the following
form of the trajectory:
\begin{equation}
\alpha^{\pm}(\sqrt{s})=a_0\pm a_1\sqrt{s}+a_2 s~,
\label{eq:n}
\end{equation}
where the sign ``$+$'' corresponds to the $N^+$ trajectory which contains 
the baryons N(939) $\frac{1}{2}^+$, N(1680) $\frac{5}{2}^+$, N(2220) 
$\frac{9}{2}^+$ and N(2700) $\frac{13}{2}^+$ \cite{PDG}, while the sign
``$-$'' corresponds to the $N^-$ trajectory, which contains 
N(1675) $\frac{5}{2}^-$ and N(2250) $\frac{9}{2}^-$ \cite{PDG}.

Thus, we obtain  $\alpha(0)= -0.398$,  
$a_1 = -(0.86\pm 1.36)\cdot 10^{-2}$ (GeV$^{-1}$) and 
$a_2 = 1.03\pm 0.01$ (GeV$^{-2}$)
 ($\chi^2/$d.o.f = 0.57). The parameter $a_0$ is eliminated from the fit by using the condition:
\begin{equation}
\alpha^{+}(m_n)\equiv a_0 - a_1 m_n +a_2 m_n^2 = 1/2~,
\end{equation}
where $m_n$ is the nucleon mass.
 
Using these values of the parameters we calculate ${\cal R}e\,\alpha^\pm (s_n)$,
where $n=1,2,x$ and $s_1=(m_{\pi}+m_N)^2=1.16$ GeV$^2$,
$s_2$ = 2.44 GeV$^2$ and $s_x$ = 11.7 GeV$^2$. In what follows, $s_2$ 
will be used as a parameter. The final result does not depend very significantly
on $s_x$, so its value is fixed at 11.7 GeV$^2$.

\noindent
Then we set $\lambda^\pm_n={\cal R}e\,\alpha^\pm(s_n)$ and continue the  
recursive fitting procedure using formulas (\ref{M5}) and (\ref{M12}).
After 10 steps the procedure converges and the result we arrive at for the
values of the parameters is
$c_1=0.22\pm 0.02$, 
$c_2=0.37\pm 0.09$, and $c_x = 18.4 \pm 1.1$, $s_2 = 2.4 \pm 0.2$ GeV$^2$, 
$\delta = 0.49 \pm 0.09$,   
($\chi^2/$d.o.f = 1.0). 
The intercept of the trajectory does not change visibly, $\alpha (0) = -0.42$.

\begin{figure}[H]
\parbox[r]{8.5cm}{\epsfxsize=75mm
\epsffile{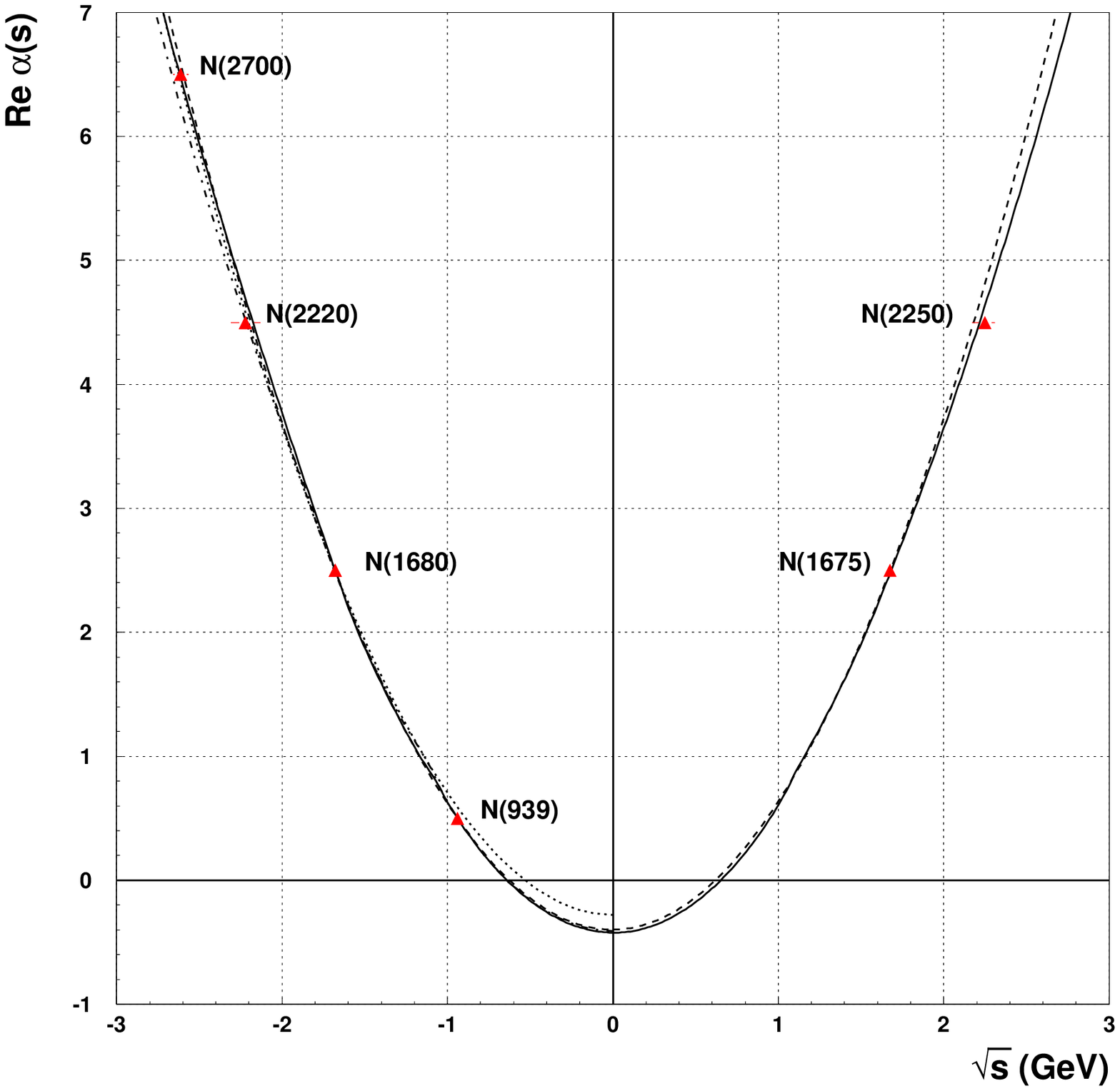}} \hfill~\parbox[c]{9.cm}{\epsfxsize=75mm
\epsffile{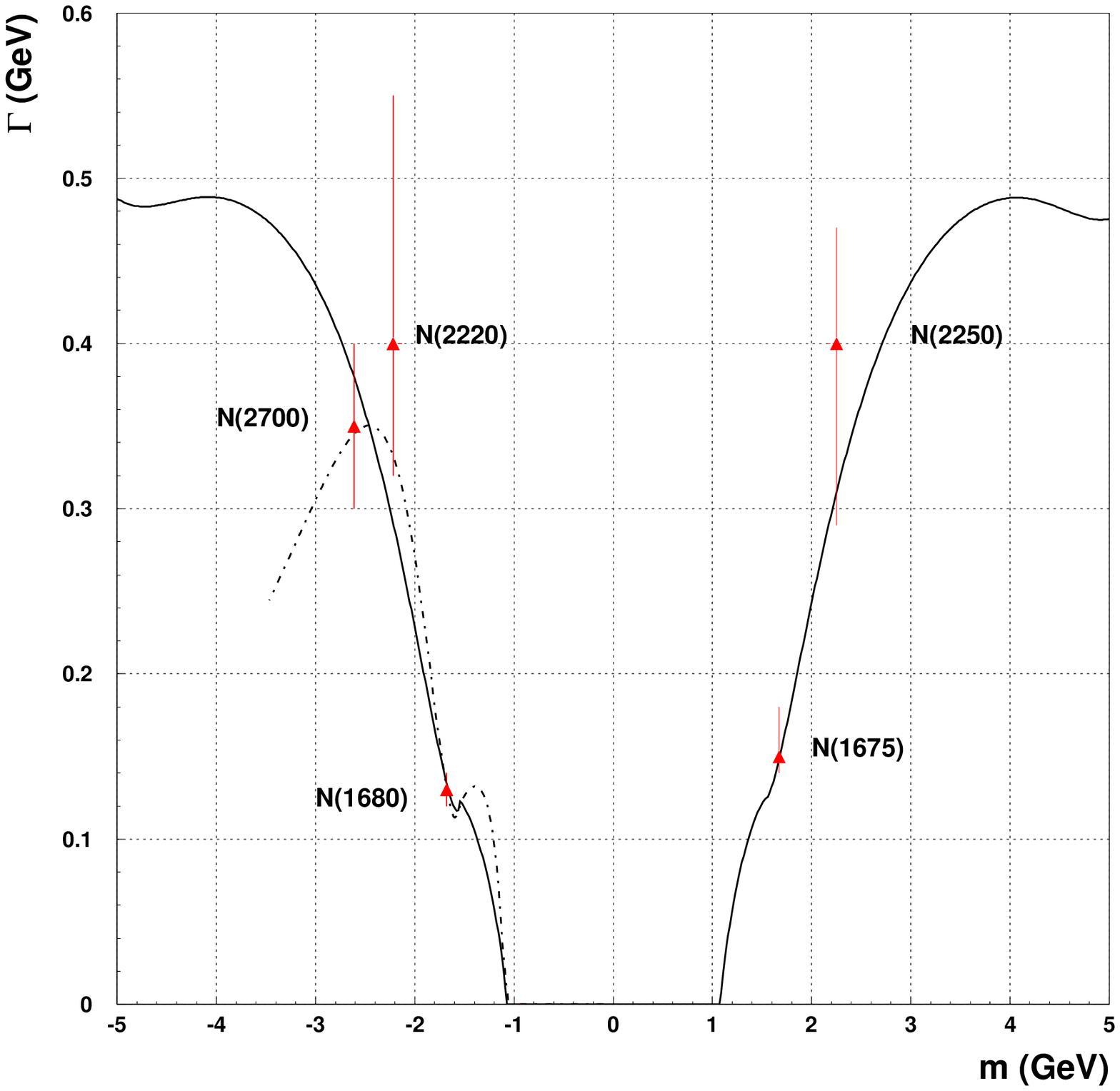}}

\parbox[t]{7.cm}{\caption{The real part of $N$ trajectory. The dashed line
corresponds to the result of the fit with Eq.~(\ref{eq:n}), the solid line 
corresponds to the final result. The dotted line corresponds to the linear fit. 
\label{fig:n}}}
\hfill~\parbox[t]{7.cm}{\caption{ The width of $N$ trajectory. 
The dotted-dashed line corresponds to the fit with the simplified model, 
the solid line corresponds to the final result.
\label{fig:nwidth}}}
\end{figure}

\noindent
The results are presented in figures~\ref{fig:n} and \ref{fig:nwidth}. 
\vskip -0.5cm

\section{Concluding remarks}

The main problem in constructing models for Regge trajectories
resides in the difficulty of making the nearly linear rise of the
real part in the whole range of observed resonances compatible
with the presence of a sizable imaginary part. Figures (1) and (3)
suggest that this difficulty has been overcome. The complexity of
the problem reveals itself in the imaginary part of the
trajectories as shown in figures (2) and (4). \vskip 0.3cm The
main assumptions regard only the imaginary part of the trajectory.
This, through dispersion relations, determines completely the
analytic structure of the model. The real part, the slope and the
widths of the resonances follow from the hypothesis that the 
contributions of different thresholds are additive. There is no
first-principle reason for this constraint, it is imposed only for
simplicity sake. \vskip 0.3cm In the simple model of Section 2 we
found a physical interpretation for the position of the thresholds
that the $N$ trajectory inherits from the partial wave
amplitude for the pion-nucleon scattering. However, such
interpretation becomes problematic when the MacDowell symmetry is
implemented. In this case, we must consider the thresholds, above
the first one, as effective thresholds. This is always true for
the last threshold $s_x$ whose position is only weakly constrained
from the experimental data. $s_x$ can move to higher values of $s$
if it is required from the discovery of new baryonic resonances.
\vskip 0.3cm Data for the excited states of the baryons are often
insufficient for a fit of the parameters appearing in the
(oversimplified) imaginary part of the trajectory. For this reason
our analysis cannot be extended to strange and charmed baryons.
Furthermore, for the nucleon, our findings do not put strong
constraints on a future search of new nucleonic states.
Approximate symmetries or dynamical models, not accounted for in
this paper, could help in obtaining more precise predictions when
complemented by analyticity and unitarity. A ``global'' fit to
baryonic trajectories could be an important step forward a deeper
understanding of their structure. This possibility will be
explored elsewhere.

\section*{Appendix A}
  \renewcommand{\theequation}{A-\arabic{equation}}
  \setcounter{equation}{0} 
In this Appendix we collect the formulas that show how 
Eq.~(\ref{M4}) can be solved and the equivalence of the solution with
the one of Eq.~(\ref{M5}). Inserting Eq.~(\ref{M6}) in Eq.~(\ref{M4})
one obtains
\begin{eqnarray}
{\cal A}_n(\sqrt{s}) & = & PV\int_0^{\infty}\,s'{}^{\delta-1/2-
\lambda_n^+}(s'-s_n)^{\lambda_n^+}\frac{1}{\sqrt{s'}-\sqrt{s}}\,d
\sqrt{s'} \theta(s'-s_n)- \nonumber \\ & - &
\int_0^{\infty}\,s'{}^{\delta-1/2-
\lambda_n^-}(s'-s_n)^{\lambda_n^-}\frac{1}{\sqrt{s'}+\sqrt{s}}\,d
\sqrt{s'} \theta(s'-s_n)~. \label{Ma7}
\end{eqnarray}
Consider now the second integral on the right hand side of 
Eq.~(\ref{Ma7}). Substituting $u=\sqrt{s_n/s'}$ we get
\begin{equation}
\int_{\sqrt{s_n}}^{\infty}\,s'{}^{\delta-1/2-
\lambda_n^-}(s'-s_n)^{\lambda_n^-}\frac{1}{\sqrt{s'}+\sqrt{s}}\,d
\sqrt{s'}=(\sqrt{s_n})^{2\delta-1}\int_0^1\,\frac{u^{-2\delta}
(1-u^2)^{\lambda_n^-}}{1+\sqrt{s/s_n}\,u}\,du~. \label{M8a}
\end{equation}
We obtain
\begin{eqnarray}
& & \int_0^1\,\frac{u^{-2\delta}
(1-u^2)^{\lambda_n^-}}{1+\sqrt{s/s_n}\,u}\,du = \nonumber \\ & &
\frac{1}{2}\Gamma(1+\lambda_n^-) \left[ -\sqrt{\frac{s}{s_n}}
\frac{\Gamma(1-\delta)}{\Gamma(2+\lambda_n^- -\delta)}
{}_2F_1\left(1,1-\delta;2-\delta+\lambda_n^-
;\frac{s}{s_n}\right)+ \right. \nonumber \\ & & \left.
+\frac{\Gamma(1/2-\delta)}{\Gamma(3/2+\lambda_n^- -\delta)}
{}_2F_1\left(1,1/2-\delta;3/2-\delta+\lambda_n^-
;\frac{s}{s_n}\right) \right]~, \label{M8b}
\end{eqnarray}
that is a convenient expression when $s_n>s$. However, when $s>s_n$,  
it is necessary to perform an analytic continuation that, in the
general case considered here \footnote{In the general case 
$1-c$, $b-a$ and $c-b-a$ in $F(a,b;c;z)$ are not integers.}, gives
\begin{eqnarray}
& & \int_0^1\,\frac{u^{-2\delta}
(1-u^2)^{\lambda_n^-}}{1+\sqrt{s/s_n}\,u}\,du = \nonumber \\ & &
\frac{1}{2}\left[ \frac{\Gamma[\lambda_n^-+1)
\Gamma(-\delta)}{\Gamma(1+\lambda_n^-
-\delta)}\sqrt{\frac{s_n}{s}}
{}_2F_1\left(1,\delta-\lambda_n^-;1+\delta;\frac{s_n}{s}\right)\right.
- \nonumber \\ & & \frac{\Gamma(1+\lambda_n^-)\Gamma(-1/2-\delta)
}{\Gamma(1/2+\lambda_n^- -\delta)}\frac{s_n}{s}
{}_2F_1\left(1,1/2+\delta-\lambda_n^-;3/2+\delta;\frac{s_n}{s}\right)+
\nonumber \\ & & \left. + \pi \left(\frac{s_n}{s}\right)^{1/2-
\delta}\left(1-\frac{s_n}{s}\right)^{\lambda_n^-}
\frac{1}{\sin(\pi\delta)\cos(\pi\delta)} \right]~. \label{M8c}
\end{eqnarray}
Eqs.~(\ref{M8b}) and (\ref{M8c}) give the result for the integral 
in Eq.~(\ref{M4}) in all possible cases. \vskip 0.3cm Of the first
integral in Eq.~(\ref{Ma7}) we must take the principal value. By
multiplying the numerator and denominator of the integrand by
$(\sqrt{s'}+\sqrt{s})$ we get
\begin{displaymath}
PV\int_0^{\infty}\,s'{}^{\delta-1/2-
\lambda_n^+}(s'-s_n)^{\lambda_n^+}\frac{1}{\sqrt{s'}-\sqrt{s}}\,d
\sqrt{s'} \theta(s'-s_n)=
\end{displaymath}
\begin{equation}
=\frac{\sqrt{s}}{2}\,PV\int_0^{\infty}\frac{y^{\lambda_n^+}(y+s_n)^{\delta-1
-\lambda_n^+}}{y-(s-s_n)}dy+
\frac{1}{2}\,PV\int_0^{\infty}\frac{y^{\lambda_n^+}(y+s_n)^{\delta-1/2
-\lambda_n^+}}{y-(s-s_n)}dy~, \label{M13}
\end{equation}
that can be easily solved since both integrals, in the second row
of Eq.~(\ref{M13}), are of the same form of the integrals solved
before. Since ${\cal R}e\,\alpha(\sqrt{s})$ is an analytic
function of $\sqrt{s}$ Eqs~(\ref{M8b}) and (\ref{M8c})
hold also when $\sqrt{s}<0$.

\section*{Appendix B}
  \renewcommand{\theequation}{B-\arabic{equation}}
  \setcounter{equation}{0} 

It is easy to show that the derivative of ${\cal
R}e\,\alpha(\sqrt{s})$ gives for ${\cal D}_n(\sqrt{s})$ the
following cumbersome expression:
\begin{eqnarray}
{\cal D}_n(\sqrt{s}) & = & \frac{1}{2} \left\{ \left[2
s^{\delta-3/2}
\left(1-\frac{s_n}{s}\right)^{\lambda_n^+-1}(\lambda_n^+s_n+\delta(s-s_n))
\tan(\pi\delta) \right. \right. \nonumber \\ & + &
\frac{s_n^{\delta+1/2}}{s}\frac{\Gamma(-1/2-\delta)\Gamma(1+\lambda_n^+)}{\pi
\Gamma(1/2-\delta+\lambda_n^+)}\left({}_2F_1\left(1,\frac{1}{2}+\delta-\lambda_n^+;
\frac{3}{2}+\delta;\frac{s_n}{s}\right) \right. \nonumber \\ & + &
\left. \frac{2(\delta-\lambda_n^+
+1/2)}{\delta+3/2}\,\frac{s_n}{s}
{}_2F_1\left(2,\frac{3}{2}+\delta-\lambda_n^+;
\frac{5}{2}+\delta;\frac{s_n}{s}\right)\right) \nonumber \\ & - &
\left. \mbox{the same term
with}\;\;\lambda_n^+ \to \lambda_n^-\vphantom{\left(\frac{s}{s_n}\right)}\right ] +  \nonumber \\
& + & \left[-2 s^{\delta-3/2}
\left(1-\frac{s_n}{s}\right)^{\lambda_n^+-1}(\lambda_n^+s_n+\delta(s-s_n))
\cot(\pi\delta) \right. \nonumber \\ & + & 2
\frac{s_n^{\delta+1}(\delta-\lambda_n^+)
\Gamma(-\delta)\Gamma(1+\lambda_n^+)}{\pi s^{3/2} (1+\delta)
\Gamma(1-\delta+\lambda_n^+)}{}_2F_1\left(2,1+\delta-\lambda_n^+;
2+\delta;\frac{s_n}{s}\right) \nonumber \\ & + & \left. \left.
\mbox{the same term with}\;\;\lambda_n^+ \to
\lambda_n^-\vphantom{\left(\frac{s}{s_n}\right)}\right ]\right\} \theta(s-s_n)  \nonumber \\
& + & \frac{1}{2} \left\{ \left[s_n^{\delta-3/2}\frac{
\Gamma(1/2-\delta)\Gamma(1+\lambda_n^+)}{\pi
\Gamma(3/2-\delta+\lambda_n^+)}\left( s_n\,\,{}_2F_1\left(1,
\frac{1}{2}-\delta;
\frac{3}{2}-\delta+\lambda_n^+;\frac{s}{s_n}\right) \right.
\right. \right. \nonumber \\ & + & \left. \frac{(1-2\delta)
s}{\lambda_n^+-\delta+3/2}{}_2F_1\left(2,\frac{3}{2}-\delta;
\frac{5}{2}-\delta+\lambda_n^+;\frac{s}{s_n}\right) \right)
\nonumber \\ & - & \left. \mbox{the same term with}\;\;\lambda_n^+
\to \lambda_n^- \vphantom{\left(\frac{s}{s_n}\right)}\right ] +  \nonumber
\\ & + & \left[2 \frac{s_n^{\delta-1}\Gamma(1-\delta)
\Gamma(1+\lambda_n^+)}{\pi\Gamma(2-\delta+\lambda_n^+)}
\sqrt{s}\,\,{}_2F_1\left(1,1-\delta;2-\delta+\lambda_n^+;\frac{s}{s_n}\right)
\right.  \nonumber
\\ & + & \left. \left. \mbox{the same term with} \;\;\lambda_n^+
\to \lambda_n^- \vphantom{\left(\frac{s}{s_n}\right)}\right ]
\right \} \theta(s_n-s)~. \label{M8d}
\end{eqnarray}

\end{document}